\documentclass[twocolumn, preprintnumbers,aps,prl,amsmath,amssymb,floatfix]{revtex4}
\usepackage{graphicx}
\usepackage{bm}

\begin{document}

\title{Evidence for orbital ordering in $LaCoO_{3}$}

\author{G. Maris$^{1}$, Y. Ren$^{2}$, V.Volotchaev$^{1}$, C. Zobel$^{3}$, T. Lorenz$^{3}$ and T.T.M. Palstra$^{1,*}$}

\address{$^{1}$Solid State Chemistry Laboratory, Materials Science Centre, University of
Groningen,\\ Nijenborg 4, 9747 AG Groningen, The Netherlands}
\address{$^{2}$Argonne National Laboratory, 9700 South Cass Avenue, Argonne, IL 60439}
\address{$^{3}$Physikalisches Institut, Universit\"{a}t zu K\"{o}ln, Z\"{u}lpicher Str. 77, 50937 K\"{o}ln, Germany}

\begin{abstract}
We present powder and single crystal X-ray diffraction data as evidence for a monoclinic distortion in the low
spin ($S=0$) and intermediate spin state ($S=1$) of $LaCoO_{3}$. The alternation of short and long bonds in the
\emph{ab} plane indicates the presence of $e_{g}$ orbital ordering induced by a cooperative Jahn-Teller
distortion. We observe an increase of the Jahn-Teller distortion with temperature in agreement with a thermally
activated behavior of the $Co^{3+}$ ions from a low-spin ground state to an intermediate-spin excited state.

\end{abstract}


\maketitle

The study of orbital degrees of freedom in transition metal (TM) oxides has gained prominent interest. Novel
techniques such as Resonant X-ray Scattering and X-Ray Absorption Spectroscopy give direct information about
orbital occupancy. It is realized that the orbital moments are as important as the spin moments to understand
the electronic properties. Prominent examples in which the orbital degrees of freedom determine the electronic
properties are the metal-insulator transitions in $V_{2}O_{3}$ \cite{Paolasini} and doped
$LaMnO_{3}$\cite{Murakami}. Here, spin- and orbital-induced transitions are intimately related. In the
perovkites, metallicity and orbital ordering are mutually exclusive because of the large Jahn-Teller (JT)
splitting of the $e_{g}$ orbitals \cite{Aken}. Furthermore, novel ground states can be obtained such as an
orbital liquid \cite{Khaliullin, Keimer}.

Considerable interest in the perovskite-type $LaCoO_{3}$ originates from the puzzling nature of two transitions
in this compound and the vicinity to a metal-insulator transition. The ground state of $LaCoO_{3}$ is a
nonmagnetic insulator and there is no long range magnetic order at all temperatures. At low temperatures, the
magnetic susceptibility increases exponentially with temperature exhibiting a maximum near 100K. At higher
temperatures, a second anomaly is observed around 500K which is accompanied by a semiconductor to metal
transition. The maximum at 100K was ascribed initially to a change of the spin state in the $Co^{3+}$ ions i.e.
a transition from a low-spin (LS) nonmagnetic ground state ($t_{2g}^{6}$, $S=0$) to a high-spin (HS) state
($t_{2g}^{4}e_{g}^{2}$, $S=2$) \cite{Asai, Itoh, Yamaguchi, Senaris}. In the more recent literature \cite{Potze,
Saitoh, Asai2, Yamaguchi2, Kobayashi}, new scenarios involving an intermediate-spin state (IS)
($t_{2g}^{5}e_{g}^{1}$, $S=1$) have been proposed. Using LDA+U calculations, Korotin \emph{et al.}
\cite{Korotin} proposed the stabilization of the IS state due to the large hybridization between the Co-$e_{g}$
and O-$2p$ levels. Due to the partially filled $e_{g}$ level, the IS state is JT active. The degeneracy of the
$e_{g}$ orbitals of $Co^{3+}$ ions in the LS state is expected to be lifted in the IS state by a JT distortion.

All structural studies, based on powder X-ray and neutron diffraction experiments are consistently interpreted
in rhombohedral $R\bar{3}c$ symmetry and no structural transitions are reported in the temperature interval
$4.2K-1248K$ \cite{Thornton, Kobayashi2, Radelli, Louca}. A cooperative JT distortion is incompatible with this
space group. The rhombohedral distortion of the parent cubic perovskite structure consists of a deformation
along the body diagonal, and preserves only one $Co-O$ distance. This triggered our present study to detect a
coherent JT distortion and to investigate why recent high quality structural studies failed to observe such
state. As the JT distortion caused by $e_{g}$ orbital ordering are usually quite large, there must be reasons
why the distortion went unobserved. Furthermore, the LS-IS transition is robust and observed both in powder and
single crystal samples. A lattice distortion due to the JT effect in the IS state has been suggested by infrared
spectroscopy measurements \cite{Yamaguchi2} and recently by magnetic susceptibility and thermal expansion
measurements \cite{Zobel}.

It is in certain cases exceedingly difficult to identify the correct crystal symmetry. In perovskites such as
$LaMnO_{3}$ or $YVO_{3}$ the JT distortion involves alternating long and short TM-O distances without always
lowering the symmetry of the lattice \cite{Blake}. Consequently, refinement of powder diffraction data are not
very sensitive to several symmetry modifications and single crystal diffraction is prerequisite to determine the
crystal symmetry. $LaCoO_{3}$ has been consistently analyzed with the rhombohedral space group $R\bar{3}c$. This
rhombohedral symmetry involves an alternating rotation of the corner sharing $CoO_{6}$ octahedra along all three
crystallographic axes of the undistorted, cubic perovskite parent structure. The corner sharing oxygen octahedra
can be described by one oxygen position. This symmetry allows only one $Co-O$ distance and thus rules out a
coherent JT distortion. Such distortion, while preserving the alternating rotation of the $CoO_{6}$ octahedra,
lowers the symmetry to monoclinic subgroups of $R\bar{3}c$. The monoclinic space group $I2/a$ \cite{cell} is a
subgroup of $R\bar{3}c$ with the highest symmetry that accommodates the rotations of the octahedra as well as a
coherent JT distortion. The symmetry reduction to $I2/a$ space group is caused by rotations of unequal magnitude
and/or a differentiation of $Co-O$ bond lengths, induced by a coherent JT distortion. In this symmetry two
oxygen positions describe the oxygen framework. A shift of the second oxygen position with $x_{O2}=y_{O2}$
corresponds with a differentiation of $Co-O$ bond lengths, whereas a movement perpendicular to this constraint
results in a coherent rotation of the octahedra.

We provide strong evidence for such monoclinic distortion in the LS/IS state using high resolution synchrotron
single crystal and powder diffraction experiments. The monoclinic distortion is ascribed to a JT distortion that
signals the long range ordering of the $e_{g}$ orbitals. Our results are in agreement with the LS/IS state
scenario in which the occupation of the IS states of the $Co^{3+}$ ions is thermally activated from the ground
LS state \cite{Zobel}.

Single crystals of $LaCoO_{3}$ were grown by the floating-zone technique in an image furnace. High resolution
X-ray diffraction measurements were performed on a triple-axis high energy diffractometer with a photon energy
of 115 keV at the beamline 11ID-C, BESSRC CAT, Argonne National Laboratory. $\omega-\chi$ scans were collected
on a single crystal of approximate dimension 1mm. Full data sets were measured on a small single crystal
fragment of approximate dimension 0.1mm using $Mo-K_{\alpha}$ radiation. The measurements were collected using a
Bruker SMART CCD camera and face indexed absorption corrections were performed using associated software
packeges\cite{Sheldrick1}. The integration of the frames was carried out using the SAINT \cite{Sheldrick1}
software incorporating the absorption correction program SADABS. The separation of the twins was done using the
program GEMINI \cite{Sheldrick1}. Structural determination and refinement was carried out using the SHELXTL
\cite{Sheldrick1} package.

We measured the temperature dependence of the magnetization of single crystal $LaCoO_{3}$. The magnetization can
be fitted well with a non magnetic ground state ($t_{2g}^{5}$) and a thermally activated $S=1$ state
($t_{2g}^{5}e_{g}^{1}$) with an activation energy of $\Delta\sim 180K$, as shown recently by Zobel \emph{et al.}
\cite{Zobel}. The stoichiometry of the crystal can be derived from the very small Curie-term, indicating less
than 1\% $S=1$ impurities. The impurity contribution gives rise to the deep minimum at 35K in the magnetic
susceptibility. The raising of this minimum in the powder samples is attributed to $Co^{2+}$ localized magnetic
moments associated with the reconstruction of the surface \cite{Senaris}. The volume of this surface is larger
for powder samples than for single crystals. Our single crystal measurement show a minimum susceptibility which
is $\sim 2$ times smaller than in the powder samples \cite{Asai2}.

A full structure determination has been made by single crystal X-ray diffraction. However the crystals are
highly twinned as is common both for orthorhombic and rhombohedral perovskites. Twinning can complicate the
symmetry determination considerably.

The rhombohedral perovskites are obtained by a compression of the parent cubic cell along one of the body
diagonals. The twinning is a result of the four body diagonals of which only one unique axis is chosen. The twin
law relates four twin fractions by $90^{\circ}$ rotation about the cubic main axes. The diffraction pattern of
our samples typically exhibited two or three rhombohedral twins. The twinning "masks" the reflection conditions
associated with the lowering of the symmetry from $R\bar{3}c$ to $I2/a$, especially the \emph{c}-glide plane and
the three-fold axis. First, the violation of the \emph{c}-glide plane will give rise to extra reflections.
However, these reflections are also generated by the rhombohedral twinning. Therefore, their presence cannot be
used as evidence for monoclinic distortion. Second, while the disappearance of the three-fold symmetry doesn't
generate new reflections, the resulting twinning mimics $R\bar{3}c$ symmetry. A small monoclinic distortion can
give rise to such pseudo-merohedral twinning, which can be easily overlooked in a standard analysis due to
insufficient angular resolution. Therefore, high angular resolution single crystal X-ray diffraction experiments
are needed to determine the symmetry.

We verified the presence of twinning by performing $\omega-\chi$ scans with high angular resolution (0.01
degrees/step) using synchrotron radiation. Such scans measure a cross section of reciprocal space on the Ewald
sphere perpendicular to the scattering vector. In this way we measure the fine structure of a Bragg peak caused
by twinning. Fig. 1 shows such a profile of a (10 2 8) reflection \cite{indices} in an $\omega-\chi$ scan
measured at 60K. The two big spots which split by 0.15(1) degrees in $\omega$, correspond to two twins generated
by lowering the symmetry from cubic to rhombohedral. Each of these rhombohedral twins shows a further splitting
of about 0.04(1) degrees in $\omega$. This smaller splitting is best observed at 20K and 60K but also the
measured rhombohedral peaks at 120K and room temperature show clear shoulders. We attribute this smaller
splitting to the twinning caused by the decrease in symmetry from rhombohedral to monoclinic. The same type of
profiles are observed for all the peaks we have measured on several crystals. Such splitting indicates a
distortion from the rhombohedral unit cell parameters. The observed splitting indicates that the monoclinic
symmetry is maintained in the temperature interval 20K - 300K.

\begin{figure}[tb]
  \includegraphics[bb= 0 0 460 415, width=6cm]{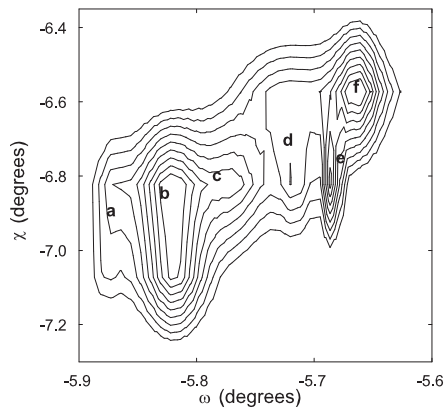}\\
  \caption{The profile of a (10 2 8) reflection \cite{indices} in an $\omega-\chi$ scan at 60K.
  The two rhombohedral fractions positioned at $w_{1} \sim -5.82^{\circ}$ and $w_{2} \sim -5.69^{\circ}$ split further into three
  peaks representing the monoclinic twins and are labelled as a, b, c and d, e, f respectively.}
\end{figure}

After establishing the symmetry of the structure, the atomic positions can be determined from standard single
crystal diffraction experiments. Full data sets were measured at different temperatures (90K, 120K, 140K, 160K,
200K, 250K and 295K) on a Bruker APEX diffractometer, using $Mo-K_{\alpha}$ radiation. The rhombohedral twinning
was clearly observable and the twins were analyzed separately. However, the monoclinic twinning could not be
observed on this diffractometer due to insufficient angular resolution. The presence of the monoclinic twins was
observed on a Enraf-Nonius CAD4 diffractometer equipped with a point detector which yields a better angular
resolution. Again, these measurements indicate the splitting of the rhombohedral peaks as evidence for
monoclinic distortion. However, for structural refinement, the monoclinic twin fractions could not be separated.
Therefore the unit cell parameters derived from the single crystal measurements are averaged at their
rhombohedral values (Table 1). The intensities of the reflections were obtained from the largest rhombohedral
twin fraction which includes the integrated intensity of the monoclinic twin fractions together. Nevertheless,
precise positions of the atoms can be obtained by making use in the refinements of the monoclinic twin relation.

The pseudo-merohedral twinning in $LaCoO_{3}$ results from the loss of three-fold symmetry during the crystal
growth or a structural phase transition at elevated temperature. Therefore, the twinning corresponds to the
three possibilities of converting the rhombohedral axes in $R\bar{3}c$ symmetry into the \emph{c}-axis in
monoclinic symmetry. There is no physical reason for one twin to be favored over the others and equal volume
fractions can be expected. All crystals that we investigated, exhibited such a monoclinic twinning. The twin law
relates three twin fractions by $120^{\circ}$ rotation about the cubic [111] direction. If the twin fractions
are equal, the reflections (h,k,l) \cite{indices} and cyclic permutations have equal intensities and simulate
the existence of a three-fold axis. Thus, it is not surprising that the quality of the refinements in both
$R\bar{3}c$ and $I2/a$ space groups are essentially the same. However, by introducing this twin model, the
quality of the refinements in $I2/a$ space group was improved , the resulting $R1$-values falling for a typical
data set (140K) from 0.0445 to 0.0433. The refined values of the twin fractions ranges between 30\% and 40\%.
The atomic coordinates for a typical data set (200K) and the selected bond lengths are displayed in Table II and
Table III respectively.

\begin{table}[htb]
\caption{Lattice parameters for $LaCoO_{3}$}
\begin{tabular}{ccccccc}
  \hline
  \hline
  T & a & b & c & $\beta$ & Volume \\
 \hline
  295 &   5.3695(4) &   5.4331(4) &   7.6395(6) &   90.985(5) &   222.83(4) \\
  250 &   5.3672(4) &   5.4329(4) &   7.6367(5) &   91.014(4) &   222.65(4) \\
  200 &   5.3611(3) &   5.4316(3) &   7.6318(4) &   91.056(3) &   222.20(3) \\
  160 &   5.3578(2) &   5.4291(2) &   7.6288(3) &   91.108(3) &   221.86(2) \\
  140 &   5.3585(2) &   5.4316(2) &   7.6304(3) &   91.126(2) &   222.04(2) \\
  120 &   5.3544(2) &   5.4294(2) &   7.6257(2) &   91.148(2) &   221.64(2) \\
  90 &   5.3469(3) &   5.4227(3) &   7.6168(4) &   91.174(3) &   220.80(2) \\
  \hline
  \hline
\end{tabular}
\end{table}
\begin{table}[htb]
\caption{Atomic parameters for $LaCoO_{3}$ at 200K}
\begin{tabular}{cccccc}
  \hline
  \hline
  Atom & Position & x & y & z & $U_{iso}$ \\
 \hline
  La &   4e &   0.25 &   0.25019(13) &   0 & 0.00486(11) \\
  Co &   4c &   0.75 &   0.25 &   0.25 & 0.00452(13) \\
  O1 &   4e &   0.25 &   -0.3068 &   0 & 0.0101(15) \\
  O2 &   8f &   0.0241(10) &   0.0332(10) &  0.2293(7) & 0.0044(6) \\
  \hline
  \hline
\end{tabular}
\end{table}
\begin{table}[htb]
\caption{Selected bond distances in \AA for $LaCoO_{3}$}
\begin{tabular}{ccccccc}
  \hline
  \hline
  Temp(K) & Co-O2(\AA) & Co-O1(\AA) & Co-O2(\AA) \\
 \hline
  295 &   1.874(7) &   1.925(8) &   1.993(8)  \\
  250 &   1.877(6) &   1.9357(9) &   1.9792(6) \\
  200 &   1.892(7) &   1.9326(9) &   1.962(7)  \\
  160 &   1.889(6) &   1.9329(9) &   1.962(6) \\
  140 &   1.882(6) &   1.9332(9) &   1.972(6)  \\
  120 &   1.879(6) &   1.9325(9) &   1.971(5)  \\
  90 &   1.918(6) &   1.9235(8) &   1.934(6)  \\
  \hline
  \hline
\end{tabular}
\end{table}

Recent high resolution powder neutron diffraction measurements were analyzed with $R\bar{3}c$ symmetry and
failed to identify a monoclinic distortion \cite{Radelli}. We performed experiments on high quality powder
obtained from a crushed single crystal using synchrotron radiation and high angular resolution: 0.001
degrees/step. We measured the temperature dependence of the profile of the (4 0 0) reflection\cite{indices} as
the degeneracy of this peak is expected to be lifted by lowering the symmetry from rhombohedral $R\bar{3}c$
space group to monoclinic $I2/a$ space group. The peak width is resolution limited ($0.0025^{\circ}$) up to 65K
and it starts to broaden gradually as the temperature is increased, reaching $0.0036^{\circ}$ at 145K. We
associate the broadening of the Bragg reflection with a gradual distortion of the structure. The monoclinic
symmetry is compatible with a splitting of this reflection. In the left side of fig. 2 we show the temperature
dependence of the splitting of the (4 0 0) reflection derived from the peak profiles. In the inset of fig. 2 we
show the profiles of the peak at 45K and at 120K. The splitting in $\omega$ derived from the $\omega-\chi$ scans
on single crystals corresponds with a much smaller splitting in $\theta-2\theta$ scans in the Bragg geometry.
Thus, due to limitation in angular resolution, the monoclinic distortion can go easily undetected in Bragg
geometry powder X-ray diffraction.

\begin{figure}[tb]
  \includegraphics[width=8cm]{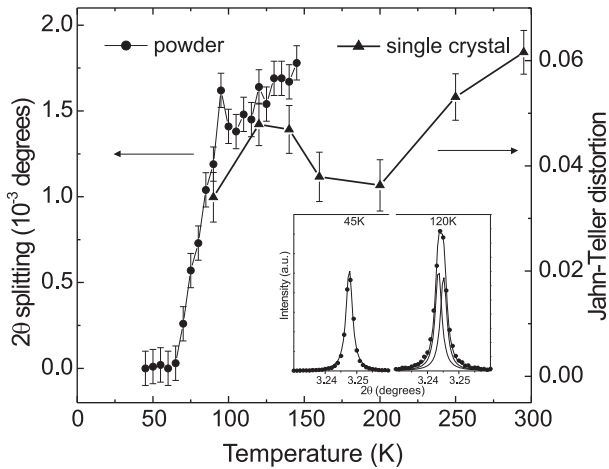}\\
  \caption{Left: Temperature dependence of the splitting of (4 0 0) reflection\cite{indices} in $\theta-2\theta$ geometry
(below 65K the line width is resolution limited). Right: The JT distortion parameter versus temperature as
determined from the single crystal refinements. The lines are guides for the eye. Inset: Profile of (4 0 0)
reflection at 45K and 120K. The profile at 120K can be accurately fitted by two Lorentzians with the line width
of the 45K data.}
\end{figure}

We notice that powder and single crystal diffraction give complementary information. The single crystal analysis
is of crucial importance for detecting the lowering of the symmetry. However, analyzing the single crystal data
turns to be difficult due to pseudo-merohedral twinning and only yield proper crystallographic results when
appropriate symmetry and twinning are used. No accurate lattice parameters can be derived from single crystal
X-ray diffraction as they are averaged by the twinning. Nevertheless, the rotation and JT distortion can be
accurately determined from a refinement. Once the correct space group is assigned, the lattice parameters can be
more precisely determined from a powder X-ray diffraction measurement.

We derive three unequal Co-O bond lengths over the studied temperature range: one short, one long and one
medium. The long and short Co-O distances correspond with the bonds in the \emph{ab} plane, while the medium
Co-O distances are the out of plane bonds. These results indicate a clear coherent Jahn-Teller effect associated
with a Q2 type of distortion \cite{Kanamori}. The right side of fig.2 shows the temperature dependence of the
Jahn-Teller parameter, defined as the difference between the long and short bond lengths normalized by their
average. The magnitude of the JT parameter increases with temperature, reaching $\sim 6\%$ at room temperature.
This value is somewhat smaller than the JT distortion of $\sim 10\%$ found in $LaMnO_{3}$ at room temperature
\cite{Rodriguez}.

The distortion of the octahedra lifts the degeneracy of the $e_{g}$ orbitals and XAS data \cite{Hao} show that
the $z^{2}$ - like orbitals are thermally populated. The $z^{2}$ - like orbitals have an antiferrodistortive
ordering along all three directions suggesting a ferromagnetic coupling between magnetic sites \cite{Korotin}.
Such a ferromagnetic exchange interaction is also supported by polarized neutron measurements \cite{Asai}. An
antiferrodistortive orbital ordering giving rise to ferromagnetic interactions was also found in $YTiO_{3}$,
where the $t_{2g}$ orbitals of $Ti^{3+}$ cations are occupied by a single electron \cite{Sawada}. However, for
this system the JT distortion involves $t_{2g}$ orbitals for which the magnitude of the distortion is roughly
five times smaller than for $e_{g}$-based JT systems.

The discovery of a coherent JT distortion in $LaCoO_{3}$ offers novel insight and is unexpected considering the
scrutiny with which this compound was studied. It is clear that the JT distortion may have gone unnoticed in
single crystal X-ray diffraction experiments because of twinning. Moreover, the JT distortion is not easily
detected by powder X-ray diffraction because of the very small monoclinic distortion.

Our measurements are in agreement with the recent magnetic and XAS measurements\cite{Zobel, Hao}, which were
interpreted with a thermal activation of the IS spin state. Considering the fact that $LaCoO_{3}$ is close to a
metallic ground state, the transition may involve more complex behavior. It is observed in the
manganites\cite{Aken, Mathur} that phases coexist of metallic and insulating states arises from a competition
between JT-distorted and non-JT distorted regions in the sample. The propagation of strain in crystal may result
in an appearance that closely resembles a transition. Such phase coexistence requires detailed real space
measurements.

We conclude that $LaCoO_{3}$ exhibits a monoclinic distorted structure in the temperature interval 20K-300K. The
monoclinic distortion state is brought about by a cooperative Jahn-Teller effect which triggers the long range
orbital ordering of the $e_{g}$ orbitals. The gradual increasing with temperature of the Jahn-Teller distortion
suggests the thermal population of the IS state of the $Co^{3+}$ ions.

We thank J.L. de Boer, M. Gr\"{u}ninger, A. Meetsma and L.H. Tjeng for stimulating discussions. This work is
supported by the Netherlands Foundation for the Fundamental Research on Matter (FOM) and by the Deutsche
Forschungsgemeinschaft through SFB 608. Use of the Advanced Photon Source was supported by the U. S. Department
of Energy, Office of Science, Office of Basic Energy Sciences, under Contract No. W-31-109-Eng-38.


\begin{references}
\bibitem{cell} The unit cell of $a \approx 9.26 \AA$, $b \approx 5.43 \AA$, $c \approx 5.37 \AA$ $\beta \approx 124.4^{\circ}$ in the standard setting $C2/c$ is transformed to $a \approx 5.37 \AA$, $b \approx 5.43 \AA$, $c \approx 7.64 \AA$ $\beta \approx 91^{\circ}$ in $I2/a$ setting.
\end{references}
\end{document}